# Martian Methyl Chloride. A lesson in uncertainty.

William Bains

Dept. of Earth, Atmospheric and Planetary Sciences, 54-1726, Massachusetts Institute of Technology, 77 Massachusetts Avenue, Cambridge, MA 02139, USA. bains@mit.edu

This paper is in support of the poster "Martian Methyl Chloride. A lesson in uncertainty." presented at UK astrobiology Conference 2013 – ARB5.

## Abstract

The MSL Lander *Curiosity* has recently detected methyl halides coming from heated samples of Martian soil. This is reminiscent of similar findings in the Viking Lander spacecraft. In the 1970s a consensus developed quickly explaining the methyl halides as contamination originating from the spacecraft, and ignoring lines of evidence that the two compounds originated from Mars, and that they could not have originated from the proposed spacecraft chemistry. I discuss why this consensus developed from the understanding of biochemistry and geochemistry of 1976, despite its implausibility. Subsequent explanations for the Viking methyl halides are more plausible but still not proven. The *Curiosity* rover results are also being explained as a result of on-spacecraft chemistry. I urge caution in this interpretation, in light of the historical Viking example: it is better to leave unexplained data unexplained than to lock in an explanation that precludes future developments.

## MSL, Viking, and methyl chloride

The MSL Curiosity Rover has recently found methyl halides released from heated soil samples on Mars. In the press conference on the results at the[1], mission team explicitly said the MSL lander "SAM [the Surface Analysis at Mars instruments] has no definitive detection to report of organic compounds with these first set of experiments", explaining that the methyl halides were a product of chemistry occurring within the lander. For those with long memories, both the detection of methyl halides and the explanations for that detection are reminiscent of the Viking missions in 1975-6, which also found methyl halides coming from heated Martian soil, and also attributed these to reactions happening in the lander. It is worthwhile revisiting the Viking data and its explanation to see what can be learned from those experiments that might inform analysis of the new MSL data. In this essay I revisit the Viking methyl chloride. An accompanying technical document I provide technical detail on the chemical kinetics results summarised here.

In 1976 the Viking mission to Mars landed two robots on Mars, with an explicit mission to search for life on the planet. Viking 1 and 2 landed on July 20th and September 3rd 1976 respectively (Soffen 1977). The mission was an outstanding technical success, but the results from the search for life were disappointingly negative, with only one experiment showing even ambiguous evidence of life (Horowitz et al. 1977; Klein 1977; Levin and Straat 1977; Oyama and Berdahl 1977; Mancinelli 1998; Klein 1999; Westamm et al. 2000).

---

[1] http://fallmeeting.agu.org/2012/events/press-conference-mars-rover-curiositys-investigations-in-gale-crater/

A fourth experiment looked for organic chemicals in the surface material of Mars. The gasses given off from heated Martian soil samples were analysed using a coupled gas chromatography – mass spectrometry (GC/MS) system (Biemann et al. 1976; Biemann et al. 1977). This was part of a broader investigation of the Martian chemistry, but also had relevance for the detection of life as living organisms would be expected to leave organic traces in the soil at levels easily detectable by the instrument. A control experiment ran the GCMS on an empty oven during the cruise phase to Mars (the Cruise Phase Blank – CPB).

The raw results were quite clear. Viking Lander 1 (VL-1) sample 1 showed a strong peak in the GC that MS identified as methyl chloride, and a number of weak peaks that were components of the synthetic solvent freon-E. VL-1 CPB showed the freon-E peaks only: there was no methyl chloride detected in the VL-1 CPB. VL-1 sample 3 (sample 2 having been lost) showed no peaks at all, probably because internal systems in the GCMS reduced sensitivity to the point where the freon-E peaks were 'invisible': however had methyl chloride been present in the VL-1 sample 3 at the same concentration as VL-1 sample 1, the expectation was that methyl chloride would have been detected (Biemann et al. 1977).

Both Viking Lander 2 (VL-2) samples showed a number of peaks of organic material, including methylene chloride. VL-2 CPB showed the same spectrum of compounds except that methylene chloride was not detected in the CPB (Biemann et al. 1977).

**Initial interpretation**

Where were the methyl chloride and methylene chloride coming from? How methyl chloride, a gas at Martian temperatures and pressures, stayed in the soil sample during collection and processing? What had made these gases, not produced by any known natural geological or biological process? The puzzlement of the team is clear in second and much more extensive report (Biemann et al. 1977), which states

> "Traces of freon-E and methyl chloride had been detected in previous tests on Earth. There is no doubt that all the freon-E is of terrestrial origin. The low level, a few tens of parts per billion, did not interfere with the experiment; in fact it provided a welcome calibration of the mass scale of the instrument. The methyl chloride, or part of it, could conceivably be indigenous to Mars. However, if it were, one would expect that ethyl chloride or methyl bromide would also be formed, but none were detected. […]. Considering all these facts, we tend to

> believe that all the methyl chloride is from terrestrial sources (chlorinated solvents or from adsorbed traces of methanol and HCl)."

In fact, there was no report of methyl chloride in the earlier, terrestrial tests on breadboard versions of the system (Biemann 1974; Biemann et al. 1976), although that does not rule out it being detected and not reported. (These were not the same instruments as were flown to Mars, and hence were not cleaned in exactly the same way as the flown instrument.) It is also unlikely that the methyl chloride was a contaminant itself as none was detected in the CPBs.

The VL-2 GC/MS gave equally odd, but different, results. Here a spectrum of organics attributed to terrestrial contamination was found in the CPB and in both Martian soil samples. However the soil samples showed methylene chloride as well. No methyl chloride was detected by VL-2. Biemann et al commented on the VL-2 results thus:

> "It is obvious that with the exception of methylene chloride all the other compounds [identified in the sample] are identical with those detected in the cruise experiment." (Biemann et al. 1977)

There was no suggestion that methylene chloride was generated by methanol and HCl: in fact, there was no explanation of the methylene chloride at all: it was tacitly assumed to be of terrestrial origin *even though* none had been detected in the CPB. The enigmatic findings of two, different, specific alkyl halides in experiments from two sites on Mars was left unexplained.

However it did not remain unexplained long.

**Subsequent firming of interpretation**

A number of possible sources for the methyl halides could be suggested – contamination from Earth, generated in the lander, of Martian geochemical origin, or made by life. In the literature, the status of methyl chloride moved rapidly from being a possible product of terrestrial contaminants

> "Considering all these facts, we tend to believe that all the methyl chloride is from terrestrial sources (chlorinated solvents or from adsorbed traces of methanol and HCl)."  (Biemann et al. 1977)

to definite contamination

> "No organic compounds, other than traces attributable to terrestrial contaminants, have been detected in soil samples analysed by the GCMS" (Mazur et al. 1978) (Committee on Planetary Biology and Evolution 1978)

and thence to known terrestrial contaminant

> "As has been reported, the result of the molecular analysis experiment of four samples of Martian surface material, two from the Viking 1 site and two from Viking 2 (one of them from underneath a rock), was a resounding 'no' (Biemann et al., 1976; 1977). At the detection limit, which was in the parts per billion range for compounds containing more than two carbon atoms and the parts per million range for compounds containing one or two carbon atoms, no organic material could be detected other than the traces of terrestrial contaminants known to be present in the two instruments." (Biemann 1979)

and thence to expected and predicted terrestrial contaminant

> "The only reduced carbon compounds found were known to be terrestrial organic contaminants such as methyl chloride, acetone … and they were *detected in the expected amounts* during the experimental runs of the Martian samples" (Margulis et al. 1979) (my italics)

> "A third possibility, the complete oxidation of organics during heating in the sample chambers, is considered by Biemann et al to be very unlikely. One argument presented is that known terrestrial organic contaminants like methyl chloride, acetone, toluene, and benzene were detected *in expected amounts* during the experimental runs on the Martian samples" (Mazur et al. 1978) (my italics)

In the process methylene chloride was grouped with methyl chloride (or ignored completely – it is hard to know which). By the late 1990s, the argument had been simplified to a clear statement that no organics had been found, which strongly implied that there was no life on Mars. For example

> "No organic compounds were detected in the Martian soil by the Viking lander GC-MS, at a limit of detectability of less than parts per billion for heavy organics and parts per million for light organics."(Mancinelli 1998)

> "The Viking gas chromatography/mass spectrometry (GC/MS) experiments found no evidence for any organic compounds of Martian origin above a few parts per billion in the upper 10 cm of surface soil, suggesting the absence of a widely distributed Martian biota"(Glavin et al. 2001)

When the experiments were replicated on very dry Antarctic soils, chlorinated compounds were again found. However these Antarctic alkyl halides also attributed to contamination, this time from the PVC bags the samples were carried in (Biemann and Lavoie 1979).

The interpretation that the alkyl halides were contamination was bolstered by the observation that they did not appear in all samples. The Martian soil was believed to be very homogenous (Clark et al. 1976; Rushnek et al. 1979), so the presence of different methyl halides at the two sites shows they could not come from the soil. (This is a circular argument – equally, the methyl halides could be evidence of soil inhomogeneity.)

Why were Biemann et al initially so hesitant to consider methyl chloride and methylene chloride as endogenous to Mars, and why did the consensus move rapidly to the definitive statement that they had been proven to be terrestrial contaminants? To an extent, lack of knowledge at the time drove the community to those conclusions.

**Background to alkyl halides in 1976**

In the mid 1970s, alkyl halides were known only as industrial chemicals. College-level textbooks (see for example (Holum 1969; Griffiths 1970; Weininger 1974; Pasto 1977)) and specialist monographs (see for example (Musgrave 1964; Hutzinger et al. 1982; Stucki et al. 1982)) emphasise the use of alkyl halides as industrials chemicals, fumigants, and their toxicity to man and to insects and fungi.

Geological sources of halogenated hydrocarbons, specifically methyl chloride and methylene chloride, were being discovered in the 1970s (Stiober et al. 1971; Jordan 2003). However they were still considered a rarity, and any plausible geochemical source would produce a mixture of methyl halides, which (as Beimann et al commented) was not what was seen.

We know now that perchlorate is a stable natural minor component of very dry soils on Mars and Earth (Hecht et al. 2009). It is likely that perchlorate would oxidize and chlorinate organic material in the Martian soil (discussed below in the Appendix). In 1976 perchlorate would not have been considered as a candidate soil component. A powerful oxidizing agent, perchlorate's principle use was as ammonium perchlorate in rocket fuel (see eg (Marsel 1959; Jacobs and Whitehead 1969)), and it was subsequently discovered to be mildly toxic (Stokstad 2005). The use of the less stable chlorate as a weed-killer and ingredient in home-made fireworks was well-known (McGregor and Jackson 1969; Oliver et al. 1972; Stavrou et al. 1978).

Biology would also be considered an implausible source for the methyl chlorides. In general, studies in the 1950s and 1960s purporting to show that life produced organohalides were either considered as revealing metabolic 'freaks of nature' or as artefactual (reviewed in (Gribble 1998)). We now know that organisms make methyl chloride, methylene chloride, and a bewildering variety of other halogenated chemicals, especially methyl halides (Lovelock 1975; Van Pee 1996; Dong et al. 2000). Thus today, if a biologist found a halocarbon in a sample suspected of harbouring life, they would be predisposed to think that the halocarbon was produced by life, particularly if only one methyl halide was produced, not a mixture (McKay 2004; Shapiro and Schulze-Makuch 2009; Dorn et al. 2011). If no other organics were found, the biologist would assume that the producing organisms were buried, and the methyl halides diffusing to the surface. In 1976, they would be strongly predisposed to think that the halocarbon was abiological, and probably industrial.

**Kinetics**

(Biemann et al. 1977) did not test whether methanol and HCl would generate methyl chloride under plausible Viking conditions. In fact, chemical knowledge of the 1970s show that it is very unlikely that methanol and hydrogen chloride present in the Viking ovens could have generated methyl chloride (discussed the Appendix.) and impossible that it could have generated methylene chloride. Recall that neither were present in the CPB samples – they *must* have been on Mars, or generated on Mars: it is highly unlikely that they were brought to Mars from Earth. The lab reaction cited as the model for the Viking data takes hours not seconds, and does not produce methyl chloride at more than 70% yield, although no unreacted methanol was detected by Viking (Buehler and Pearson 1970; Dorn 1974; Furniss et al. 1978). The lab reaction is with liquid methanol. The gas phase reaction (which has been studied extensively (Thyagarajan et al. 1966) would take an hour to produce the observed results, even if reacted over an efficient, industrial catalyst. Methylene chloride can be generated from organics, chlorine and oxidizing agents via oxychlorination (Price 1957; B. F. Goodrich Company 1965; Leduc and Dubois 2008), although these reactions generate a mixture of methyl halides. 25 years after the Viking data (Navarro-Gonzalez et al. 2010) tested the ability of Atacama desert soils dosed with perchlorate to generate methyl halides through oxychlorination, and showed that they could do so, but only at temperature above those at which methyl halides were evolved by Viking. The experience of (Navarro-Gonzalez et al. 2010), and of industrial processes for generating methyl halides, suggests that oxychlorination is not impossible as a source of the methyl halides: whether it is more plausible than other explanations depends on one's view of how plausible the other explanations are.

**Other reactions**.

At least the methanol chlorination and oxychlorination reactions are thermodynamically feasible. Explanations for methyl halides at the time and since have also included reaction of carbon dioxide or carbon monoxide from the atmosphere with hydrogen chloride or chlorine. Table 1 lists the thermodynamics of these reactions, suggesting that none are plausible sources of methyl halides.

**Conclusion**

Chemistry continues to surprise us. In 1976 the perchlorate in the Martian soil was unknown (and implausible), the possibility that geochemistry and life could be a source of methyl halides was unknown, and so only one explanation for the methyl halides found by Viking was considered. Because it was the only one considered, it was not explored, and the kinetic implausibility of the mechanism was not uncovered.

The Viking mission was a technical triumph, and these are minor flaws in interpretation of a flood of data from an astonishing achievement. Today's missions to Mars continue to amaze. I hope that this discussion is a contribution to keeping the interpretation of the new data from MSL open to continued enquiry and experimentation. It is better to leave unexplained data unexplained than to lock in an explanation that precludes future discoveries.

**Acknowledgements**

An earlier version of this draft was rejected vigorously by two referees, whose conflicting opinions on what was wrong with it have, I hope, directed me to make it a bit better. I am grateful to them both: they know who they are. This is an unpublished paper and, while several colleagues have made encouraging noises about it, it has not been refereed in its current form.

**Table 1: other chloromethane-generating reactions**

| Reaction | $\Delta G^o$ Free energy (kJ/mol) | |
|---|---|---|
| | 200°C | 500°C |
| $CO_2 + 6HCl \rightarrow CH_2Cl_2 + 2H_2O + 2Cl_2$ | 489 | 565 |
| $CO_2 + Cl_2 + H_2O \rightarrow CH_2Cl_2 + \frac{3}{2} O_2$ | 564 | 579 |
| $CO + Cl_2 + H_2O \rightarrow CH_2Cl_2 + O_2$ | 324 | 366 |
| $CO + 4HCl \rightarrow CH_2Cl_2 + H_2O + Cl_2$ | 275 | 357 |

Reactions that could generate methylene chloride from carbon dioxide or carbon monoxide. Thermodynamic data from (Chase 1998). Note that positive values of $\Delta G^o$ imply that the reaction requires energy input to make it happen, i.e. the reaction will not happen spontaneously.

APPENDIX

**The Viking Volatile Release Experiment**

Samples of the loose surface material were scooped up, ground up to reduce grain size, dropped into a small oven, and heated in steps. Raising the temperature took 1 – 8 seconds, and the oven was then held at that temperature for 30 seconds(Biemann et al. 1976; Biemann et al. 1977). Any gas given off was analysed by the GC/MS. An effluent divider vented high fluxes of GC output to the Martian atmosphere to prevent the mass spectrometer being over-loaded. The operation of the effluent divider meant that the sensitivity varied between experiments. Both Viking landers had the capability to analyse 3 samples, but on both landers technical failures in one oven meant that only two samples were analysed. The oven and GC/MS system was tested by heating one oven and analysing the gases given off from the empty oven while the spacecraft were en route to Mars, to provide a "cruise phase blank" (CPB) sample.

Methyl halides were not detected in either CPB test. Thus their detection on Mars **must** be due to a factor related to Mars. The interpretation published at the time was that the alkyl halides were generated in the ovens by reaction of methanol and hydrogen chloride.

**Kinetics of methyl chloride formation**

The textbook reaction

$$CH_3OH + HCl \rightarrow CH_3Cl + H_2O \qquad (1)$$

is a specific example of conversion of an alcohol to an alkyl halide

$$ROH + HX \rightarrow RX + H_2O \qquad (2)$$

(see eg (Furniss et al. 1978) – the majority of the references below are contemporary with the Viking mission, to give a picture of the state of the art at the time: modern textbooks do not differ significantly.) However this textbook reaction is less applicable to the Viking results than it may appear.

The textbook reaction applies to passing gaseous HCl over or through boiling alcohol in the presence of a zinc chloride: the zinc chloride acts both as a catalyst and as a dehydrating agent. The rate of the general reaction above depends on the halide (reactivity in the order HI > HBr > HCl) and the alcohol (reactivity in the order tertiary > secondary > primary). Under laboratory conditions, reactions between HCl and a primary alcohol such as methanol that are run for several hours typically yield 60% - 75% product, ie 25% of the methanol remains unconverted (Buehler and Pearson 1970; Dorn 1974; Furniss et al. 1978). For this reason, preparation of alkyl halides is usually done using acid halides as a halogen donor, which reaction proceeds faster and to completion (Weininger 1974; Furniss et al. 1978). Even if a conversion efficiency of 70% had been reached in the 30 seconds reaction time in Viking (extremely unlikely in light of lab experience), the 6ppb unreacted methanol left should have been detected by the Viking GCMS instrument: none was detected.

The text-book laboratory process is a poor model for the chemistry in Viking. The Viking chemistry did not occur in solution, but in gas phase. Gas phase reaction has been used industrially to prepare methyl chloride, and its kinetics have been studied. The industrial reaction is catalysed by $ZnCl_2$/alumina or $CdCl_2$/alumina catalysts (Thyagarajan et al. 1966). From (Thyagarajan et al. 1966) the reaction rate $r$ for reaction (2) with an excess of optimal γ-alumina catalyst is given by

$$r = k[C_M].[C_H] \text{ mol}^. \text{sec}^{-1}$$

where $[C_M]$ and $[C_H]$ are the concentration of methanol and hydrogen chloride respectively in atmospheres, and $k$ is the rate constant given by

$$k = 6984 \; .e^{-\frac{19,178}{1.986 \; .T}} \text{ mol.sec}^{-1}.\text{atm}^{-2}$$

where T is the absolute temperature. To generate 15ppb $CH_3Cl$, there must be at least 15ppb HCl in the starting material. Assuming 20ppb HCl (none was detected in any Viking experiment, so there cannot have been a large excess), the time needed to generate 15ppb methyl chloride can be calculated for a range of different methanol concentrations. The result of this calculation is shown in Figure 1.

Two sets of conditions are compatible with finding 15ppb methyl chloride in the effluent gases from this reaction.
  i)   a low concentration of methanol (<200ppb) must be incubated with HCl at 200°C for at least $10^3$ seconds to generate 15ppb methyl chloride, far longer than the Viking oven heating time
  ii)  a higher concentration of methanol (>2000 ppb) incubated for 30 seconds will generate 15ppb methyl chloride, but leave behind a very large excess of unreacted methanol, which was not detected by Viking.

Thus for the proposed reaction to generate methyl chloride in the 30 seconds available, there would have to be a vast excess of methanol, which would have been detected in the instrument.

This assumes that the Martian soil is as efficient a catalyst of this reaction as an industrial alumina/chloride catalyst. If the soil is less efficient a catalyst, then the reaction is even less likely to yield the observed results.

**Side-reactions and undetected dimethyl ether**

In addition, methanol (if it had been present) would have undergone other reactions. It is well known that methanol passed over a variety of alkali catalysis such as silica and alumina will dehydrate in an exothermic reaction to form dimethyl ether (Adkins and Perkins 1928). The rate of methanol dehydration are significant over catalysts at 200$^o$C, although the rate is very slow in the absence of solid catalyst (Figueras et al. 1971). The kinetics of the reaction are more complicated than those of the methyl chloride formation above, as the rate-limiting steps depend on multi-step chemistry at the catalyst surface and hence depend sensitively on catalyst mass, particle size and other parameters (Figueras et al. 1971). However as a rough approximation, the reaction is second order in methanol in the absence of significant water – see equation 1 of (Bercic and Levec 1993).

Figure 2 shows an estimate of the time needed to dehydrate methanol (in the *absence* of competing reactions from hydrogen chloride). While this is a rough estimate only, it does suggest that the time needed to produce methyl chloride in detectable amounts is similar or greater than the time needed to produce dimethyl ether in detectable amounts, and so if methanol had been present it would have also generated dimethyl ether. No dimethyl ether was detected above the 10 ppb level.

(I note that industrial production of methyl chloride from methanol generates very little dimethyl ether. This is because the concentration of hydrogen chloride is far higher in the industrial process, favouring methyl chloride production.)

Methanol can react with methyl chloride to form dimethyl ether, an additional source of this reaction product (Packer and Vaughan 1958), and methyl chloride can react with water or base to reform methanol (Fieser and Fieser 1950). Whether these reactions will occur at a rate comparable to the formation of methyl chloride or dimethyl ether depends on the catalytic surfaces available, soil pH and other imponderables. However they can only have made the methyl chloride less pure if they occurred at all.

Thus the formation of methyl chloride from methanol and hydrogen chloride seems doubly implausible. It requires a reaction rate that is not compatible with the known gas-phase catalysed reaction of methanol and hydrogen chloride, and no dimethyl ether side-product was generated when at least some would be expected.

**Methylene chloride**

It is obviously impossible for reaction (1) to generate the methylene chloride that was detected in both VL-2 samples. The absence of methylene chloride from any Viking 1 sample and from the Viking 2 CPB very strongly suggests that it was the product of Mars, not contamination from methylene chloride itself brought from Earth. In principle, methanol could have been oxidised by the Martian regolith (which Viking discovered to be highly oxidizing (Klein 1978; Mazur et al. 1978; Klein 1979; Oro and Holzer 1979)), and then the resulting formaldehyde could then be hydrochlorinated thus:

$CH_3OH + 2[O] \rightarrow CH_2O + H_2O$ (3a)

$CH_2O + 2HCl \rightarrow CH_2Cl_2 + H_2O$ (3b)

If chlorine gas was present in the evolved gases, then direct chlorination of methane or methanol is possible

$CH_4 + 2Cl_2 \rightarrow CH_2Cl_2 + 2HCl$ (3c)

$CH_3OH + Cl_2 \rightarrow CH_2Cl_2 + H_2O$ (3d)

(Chlorine could be generated by the Deacon Process (Lewis 1906),

$4\ HCl + O_2 \rightarrow 2Cl_2 + 2H_2O$ (4)

so these are closely related reactions: in the light of later discovery, we now know that chlorine could also have been generated by breakdown of perchlorate (Navarro-Gonzalez et al. 2010)).

Combined oxidation and chlorination reactions are well known as an industrial route for generating mixed methyl halides from methane (although the complex and ill-defined chemistry involved means that these are not routes favoured for laboratory synthesis).

Yields are typically 30 – 40% of chlororinated methanes, with CO, $CO_2$ being the only major other products (B. F. Goodrich Company 1965). $CO_2$ and CO would be ignored by the analysis as being normal components of the atmosphere, so in the context of the Viking experiments the only product detected would be methylated hydrocarbons. Reactions described in (B. F. Goodrich Company 1965) typically are run at 200°C - 300°C at 1 – 5 bar pressure (Price 1957; B. F. Goodrich Company 1965; Leduc and Dubois 2008) for seconds to minutes: it is not clear if the conditions of the Viking ovens could have produced significant methylene chloride *even if* the Martian regolith was as efficient a catalyst of this reaction as industrially optimized catalysts are. Again, therefore, we suspect that the chemical kinetics do not favour the reaction.

25 years later after the Viking mission, (Navarro-Gonzalez et al. 2010) demonstrated that organics in a soil from the Atacama desert could react with perchlorate on heating, to generate methyl halides. The Atacama soils are potentially an analogue of the Martian regolith, containing perchlorate, but are known also to contain low levels of biogenic organics. (Biemann and Bada 2011) claims that (Navarro-Gonzalez et al. 2010)'s reaction conditions are different from those in Viking, but does not argue why this would change the quantitative results.

(Navarro-Gonzalez et al. 2010)'s results showed evolution of methyl chloride and methylene chloride, in soils to which perchlorate had been added, with very low yields at temperatures below 500°C. None was evolved at the 200°C shown to generate methyl chloride in VL-1 sample 1, nor at the 350°C that generated methylene chloride from both samples in VL-2. Benzene and toluene were also generated from unmodified soil, but these were oxidised by added perchlorate.

(Navarro-Gonzalez et al. 2010) also put methanol and HCl into their pyrolysis system to see if they could use the reaction proposed by (Biemann et al. 1977) to generate methyl chloride. They used 0.4% HCl in methanol at temperatures of 500°C, and as predicted from Figure 1, at these elevated temperatures and high concentrations of reagents they did see some methyl chloride produced, but, also as expected from the chemistry, saw a variety of other compounds produced.

It is notable, however, that oxychlorination reactions produce methyl chloride *and* methylene chloride. The ratio of these two products can vary depending on conditions and feedstocks, but it is highly improbable that a set of conditions can be found that would generate only methyl chloride in one lander and only methylene chloride in the other without postulating substantial differences in the regolith chemistry between the two Viking sites. Many conditions would also be expected to generate some trichloromethane (chloroform), which was not detected in any Viking experiment. (Navarro-Gonzalez et al. 2010) model the kinetics of the reaction and show that it can be biased strongly towards methyl chloride or methylene chloride, depending on the perchlorate concentration: however this model assumes that the carbon in the methyl halides is derived from methane, which is not the starting material in their actual samples (Biemann and Bada 2011). Their experimental results (which are consistent with the industrial process in producing both methyl chloride and methylene chloride) are better models for the proposed Martian reactions.

Thus while oxychlorination remains a more plausible mechanism for generating methyl chloride and methylene chloride than reaction of methanol and hydrogen chloride, neither (Navarro-Gonzalez et al. 2010)'s results not the industrial processes cited above give confidence that relatively pure materials could be generated in the time and at the temperature relevant to Viking.

The idea that the methyl chloride and methylene chloride detected in VL-1 and VL-2 respectively were generated from terrestrial contaminants in the Viking instrument therefore seems implausible. The idea that they could have been generated from reaction of organics and perchlorate in the Martian regolith is chemically more reasonable, but the observed (as opposed to theoretical) reaction kinetics (Navarro-Gonzalez et al. 2010) do not fit with those seen in the Viking landers at 200°C .

**Figure 1: Time to generate methyl chloride**

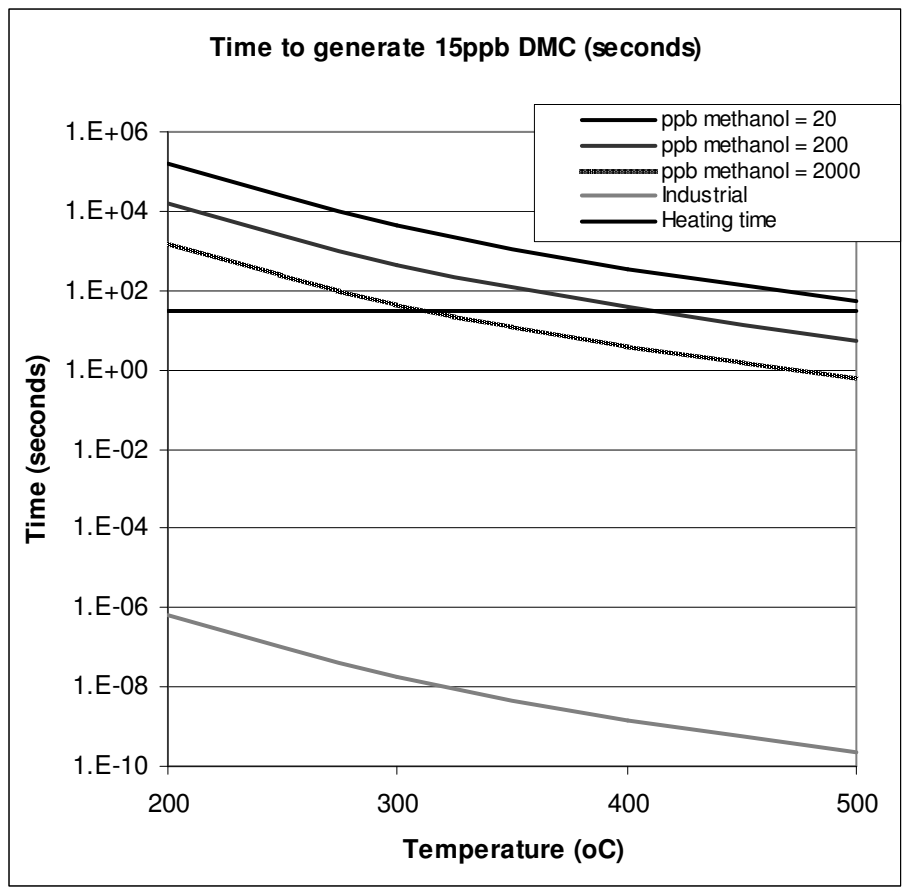

Time needed to generate 15ppb methyl chloride for different starting concentrations of methanol, assuming 20 ppb hydrogen chloride. Kinetic constants from (Thyagarajan et al. 1966). X axis: Temperature ($^oC$). Y axis: time (seconds). The four lines are for an initial concentration of methanol of 20ppb, 200ppb, 2000ppb, and 1% ("industrial"). 1% methanol in the Viking oven will generate a vapour pressure of 2.8 atmospheres, roughly corresponding to industrial high-pressure synthesis of methyl chloride from methanol and hydrogen chloride gas over an alumina catalyst. Note that one day contain 86400 seconds. Horizontal bar = 30 seconds

**Figure 2: Kinetics of dehydration of methanol**

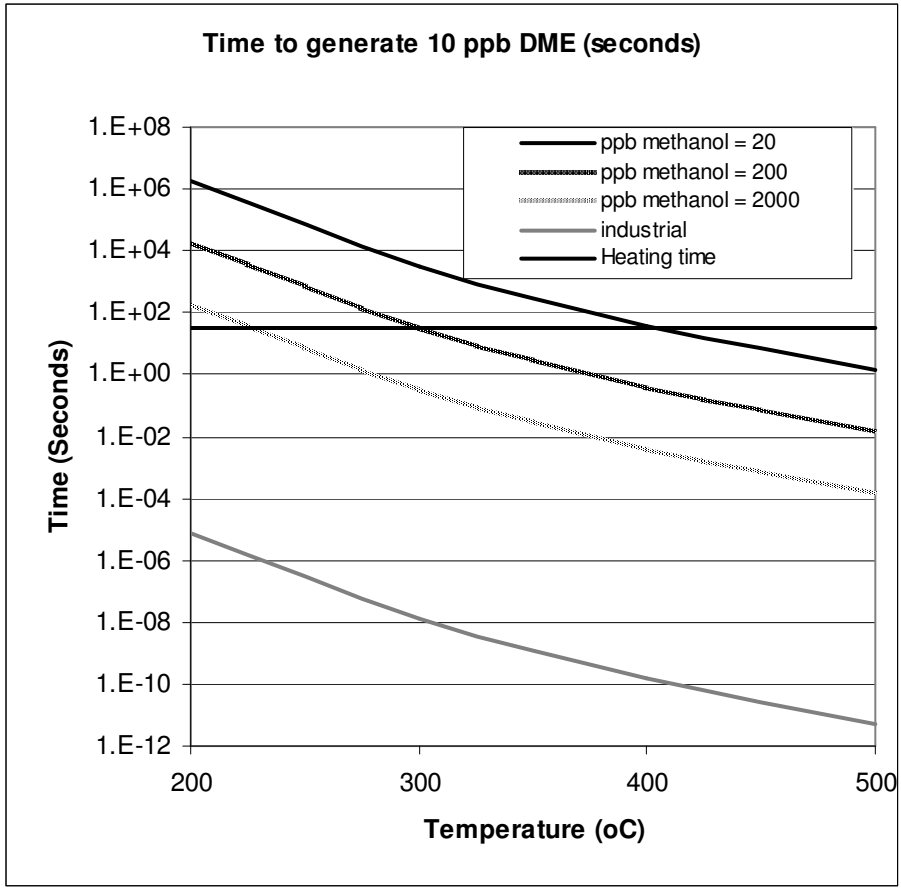

The time necessary to generate 10ppb dimethyl ether (in seconds, Y axis) vs reaction temperature (X axis) for the same ranged of methanol concentrations shown in Figure 1. Kinetic equations are from (Figueras et al. 1971; Bercic and Levec 1993). Horizontal bar = 30 seconds